\documentclass[aps,pra,twocolumn]{revtex4}
\usepackage{amsfonts}
\usepackage{amssymb}
\usepackage{color}
\usepackage{amsmath}
\usepackage{graphicx}
\usepackage{subfigure}
\usepackage{txfonts}
\usepackage{ulem}
\usepackage[colorlinks,linkcolor=blue,citecolor=blue]{hyperref}

\newcommand{\jj}[1]{{\color{black}#1}}
\newcommand{\tss}{\text{ss}}

\begin{document}

\title{Crossover from discontinuous to continuous phase transition in a dissipative spin system with collective decay}

\author{Linyu Song and Jiasen Jin}
\email{jsjin@dlut.edu.cn}
\affiliation{School of Physics, Dalian University of Technology, 116024 Dalian, China}

\begin{abstract}
We investigate the steady-state phase transitions in an all-to-all transverse-field Ising model subjected to an environment. The considered model is composed of two ingredient Hamiltonians. The orientation of the external field, which is perpendicular to the spin interaction, can be tuned to be along either $x$ direction or $z$ direction in each ingredient Hamiltonian while the dissipations always tend to flip the spins down to the $z$ direction. By means of mean-field approximation, we find that the quasicontinuous steady-state phase transition is presented as a consequence of the merging of two branches of steady-state solutions. The emergence of bistability is confirmed by analyzing the steady-state behaviors of a set of finite-size systems which is also revealed by the Liouvillian spectrum.

\end{abstract}
\date{\today}
\maketitle

\section{Introduction}
Any realistic system should be considered as an open system coupled to the external environment which influences it in a non-negligible way \cite{breuer2002}. Due to the exchange of energy or particles between the system and its environment, the state of the system will be driven away from equilibrium in the long-time limit accompanied by a breaking of detailed balance among the microstates \cite{henkel2008}. Through tuning the controllable variable of the system the ordering may appear in the steady states of the open systems. The nonequilibrium features in the steady states of open systems are observed in diverse situations ranging from the collective behaviors of creatures such as the flocking, schooling of fish, to the evolution of financial market and traffic models.

Recently, the steady-state properties of open systems in quantum domain have attracted increasing attention. The competition between the coherent evolution governed by the Hamiltonian of the system and non-unitary dissipative process induced by the system-environment interaction leads to rather rich phenomena in the long-time steady state. For instance, the exotic steady-state phases are presented in the open quantum many-body system which do not have counterparts in the closed systems \cite{lee2011,lee2013,schiro2016,rota2019,landa2020a,xl2021a,xl2021b}. Moreover, the limit cycle, which is time-dependent steady state with stable period, may also appear in the long time limit \cite{ludwig2013,jin2013,iemini2018,tucker2018,landa2020,carollo2022}, providing a new route to the intriguing time crystal in which the continuous time-translational invariance is broken \cite{sacha2018}.

Thanks to the recent experimental progress, the open quantum many-body system can be studied in the platforms of the ensemble of Rydbergs atoms \cite{carr2013,gutierrez2017,ding2020}, system of trapped ions \cite{muller2012} and array of superconductive resonant cavities \cite{fitzpatrick2017,collodo2019,ma2019}. The steady-state properties of the open quantum many-body system as well as the dynamical behavior during the time-evolution have shown to be of promising applications in the quantum state engineering \cite{plenio1999,verstraete2009,marcos2012,kouzelis2020}, quantum sensing \cite{raghunandan2018,ding2022} and simulations of the epidemic dynamics \cite{cpe2017,wintermantel2021}.

Generally the dynamics of an open quantum system can be extracted from the unitary dynamics of the joint system (system + environment) by averaging over the effects of the environmental states on the system of interest. A commonly used theoretical description of the time-evolution of the system state is the memoryless quantum master equation in the Lindblad form under the Born-Markovian approximation. However the complexity of solving the Lindblad quantum master equation scales exponentially as the system size increasing, analytical solutions can be obtained only for a few boundary-driven models \cite{prosen2011,popkov2020}. Several powerful numerical methods for simulating the dynamics of open quantum many-body system are developed and advances are made in recent years, including the corner space renormalization method \cite{finazzi2015,rota2017b}, tensor network method \cite{cui2015,mascarenhas2015,werner2016,gangat2017,kshetrimayum2017}, variational method \cite{weimer2015a,weimer2015b}, neural-network approach \cite{yoshioka2019,nagy2019,hartmann2019,vicentini2019,liu2022}, and (discrete) truncated Wigner approximation \cite{carusotto2005,vicentini2018,schachenmayer2015,huber2021,singh2022,huber2022,mink2022}. On the other hand, the Gutzwiller mean-field (MF) theory can decouple the many-body quantum master equation into single-site one by factorizing the total density matrix of the system into tensor product of identical density matrices of each site. The MF approximation can already unveil the exotic physics in the open quantum many-body systems and is shown to become accurate for high dimensional systems \cite{jin2016,huybrechts2020}. A comprehensive review on the simulation methods can be found in Ref. \cite{weimer2021}.

One of the subjects of particular interests is the critical behavior of the open quantum many-body system at the vicinity of the steady-state phase transition, such as the critical slowing down of the dynamics \cite{vicentini2018,rota2018}, the order of phase transition and the critical exponent \cite{rota2017b,minganti2018,jin2018,jin2021}. It is interesting that  jump operators that characterizes the specific dissipations may play essential role in determining the orders of the steady-state phase transition even if sharing the same Hamiltonian. For instance, the cluster mean-field results predict that the dissipative transverse-field Ising model on two-dimensional square lattice with nearest-neighboring interaction exhibits a first-order (discontinuous) phase transition when the dissipation acts along the interaction direction, while a second-order (continuous) phase transition when the dissipation acts along the transverse field \cite{jin2018}. Motivated by such a specific model, a nature question arises that how the orders of phase transition being influenced by the orientation of the dissipation.

In this paper we will demonstrate the effects of the orientation of the incoherent dissipation on the steady-state phase transition of a full connected transverse-field Ising model. In practice, for convenient we consider the collective dissipation that tends to flip the spins down to the $z$-orientation and a mixed Hamiltonian that composed of two ingredients. The proportion of the two ingredients in the total Hamiltonian can be tuned by a controlling parameter to realize the continuous changing the relative directions between the interaction (or the transverse-field) and the dissipation. We restrict the discussion in the Dicke states, by using both the MF approximation and the full quantum analysis, we provide insight into the mechanism of the crossover from the discontinuous to continuous steady-state phase transitions.

The paper is organized as follows. In Sec. \ref{sec_model}, we introduce the considered model and the quantum master equation that describes the time-evolution of the system state. We then derive the Bloch equation within the mean-field approximation and the associated Jacobian for stability analysis. In Sec. \ref{sec_results}, we first discuss the steady-state phase transitions in two limit cases in which both discontinuous and continuous phase transitions are present. We then interpolate these two limit cases by tuning the controlling parameter to investigate the effects of the orientations of the external fields (as well as the spin interactions) and the dissipation on the continuity of steady-state magnetization. We summarize in Sec. \ref{sec_summary}.

\section{The Model}
\label{sec_model}
We consider an ensemble of $N$ spin-1/2 particles with the $m$-th spin characterized by the spin angular momentum $\hat{\textbf{J}}_m = \{\hat{J}_{x,m},\hat{J}_{y,m},\hat{J}_{z,m}\}=\frac{\hbar}{2}\{\hat{\sigma}_{x,m},\hat{\sigma}_{y,m},\hat{\sigma}_{z,m}\}$ where $\hat{\sigma}^\alpha$ with $\alpha=x,y,z$ are the Pauli matrices. The spins interact to each other along $x$-direction and/or the $z$-direction via the Ising-type interactions and, at the mean time, are driven by external fields along $z$-direction and/or the $x$-direction. The Hamiltonian is given by (set $\hbar=1$ hereinafter)
\begin{equation}
\hat{H} = (1-p)\hat{H}_0 + p\hat{H}_1,
\label{Ham}
\end{equation}
where $0\le p\le1$ is a real parameter that controls the proportion of the Hamiltonians $\hat{H}_0=\frac{V}{2N}\hat{J}_x^2+g\hat{J}_z$ and $\hat{H}_1=\frac{V}{2N}\hat{J}_z^2+g\hat{J}_x$. We have introduced the collective operators $\hat{J}_\alpha=\sum_{m}{\hat{J}_{\alpha,m}}$ ($\alpha=x,y,z$). The parameter $V$ is the strengths of the spin interactions, $g$ is Rabi frequency of the external fields and $N$ is the number of spins in the system. Both $\hat{H}_0$ and $\hat{H}_1$ describe the all-to-all transverse-field Ising models and are related to each other by the transformation $\{\sigma_x,\sigma_y,\sigma_z\}\rightarrow\{\sigma_z,\sigma_y,\sigma_x\}$.

If the collective loss of one excitation is considered, the time-evolution of the state of the system can be described by the following quantum master equation
\begin{equation}
\frac{\text{d}}{\text{d}t}\hat{\rho}=\mathcal{L}\hat{\rho}=-i[\hat{H},\hat{\rho}] + \frac{\Gamma}{2N}\left(2\hat{J_-}\hat{\rho}\hat{J}_+-\{\hat{J}_+\hat{J}_-,\hat{\rho}\}\right),
\label{QME}
\end{equation}
where $\hat{\rho}$ is the density matrix of the system, $\mathcal{L}$ is the Liouvillian superoperator, $\Gamma$ is the decay rate and $\hat{J}_\pm=\hat{J}_x\pm i\hat{J}_y$ is the jump operator that tends to flip the spin down to the $z$-direction. The notation $\{\cdot,\cdot\}$ stands for the anticommutator. One can see that the orientation of dissipation can be set to parallel to the external field ($p=0$) or to the interaction ($p=1$). Actually, the Hamiltonian in Eq. (\ref{Ham}) can be recast into the following Lipkin-Meshkov-Glick (LMG) Hamiltonian \cite{morrison2008prl,morrison2008pra},
\begin{equation}
\hat{H}\propto p\hat{J}^2_x + (1-p)\hat{J}^2_z + \hat{H}_{\text{driving}},
\label{TAT_Hamiltonian}
\end{equation}
where $\hat{H}_{\text{driving}}$ is an effective driving field.

\jj{Eq. (\ref{TAT_Hamiltonian}) is a two-axis twisting (TAT) Hamiltonian for generating the squeezed spin states which are of significant applications in quantum information processing and quantum metrology. The TAT-type Hamiltonian can be implemented through imposing Rabi pulse sequences or continuous driving fields on the one-axis twisting (OAT) Hamiltonian \cite{liu2010,huang2015}. The OAT-type Hamiltonian has been realized in the BEC atomic ensembles \cite{leroux2010,riedel2010,gross2010} and the transverse and longitudinal fields can be tuned independently \cite{gil2014}. Although the experimental realization of TAT-type Hamiltonian (namely, the LMG Hamiltonian in Eq.  (\ref{TAT_Hamiltonian})) is still challenging, it is shown to be feasible to engineer the full connected LMG model in Rydberg atoms \cite{borregaard2017,nguyen2018} and atomic ensembles with an additional cavity mode \cite{groszkowski2022,zeyangli2022}. The latter also allows one to investigate the collective dynamics of dissipative systems.}

\subsection{Dicke states}
Define the total angular momentum operator $\hat{J}^2=\hat{J}_x^2+\hat{J}_y^2+\hat{J}_z^2$, one has the usual angular momentum commutation relations
\begin{eqnarray}
[\hat{J}^2,\hat{J}_\alpha]&=&0, \forall\alpha=x,y,z\cr\cr
[\hat{J}_z,\hat{J}_{\pm}]&=&\pm\hat{J}_{\pm}.
\end{eqnarray}
Then we introduce the simultaneous eigenstates $\{|j,m\rangle\}$ of $\hat{J}^2$ and $\hat{J}_z$ which satisfy
\begin{eqnarray}
\hat{J}^2|j,m\rangle&=&j(j+1)|j,m\rangle,\cr\cr
\hat{J}_z|j,m\rangle&=&m|j,m\rangle,
\end{eqnarray}
where $j\le\frac{N}{2}$ is an integer or half-integer denoting the quantum number of the total angular momentum for even or odd number of $N$ and $|m|\le j$. Further, the collective jump operator $\hat{J}_-$ and its Hermitian conjugation $\hat{J}_+$  act on $|j,m\rangle$ in the following way
\begin{equation}
\hat{J}_\pm|j,m\rangle=\sqrt{j(j+1)-m(m\pm1)}|j,m\pm1\rangle.
\end{equation}

In particular, the eigenstate $|j,m\rangle$ for the maximum $j=\frac{N}{2}$ is referred to as the Dicke state. Since $\hat{J}^2$ commutes with all all components of the angular momentum $\hat{J}_\alpha$, the total angular momentum $j$ is preserved during the time-evolution governed by Eq. (\ref{QME}). Therefore if the state of the system is initialized in one of the Dicke states $|\frac{N}{2},m\rangle$, the collective process reduces the $2^N$-dimension Hilbert space into a $(N+1)$-dimension state space. We note that if the local dissipation (e.g. the spontaneous emission of each spin) is considered, the neighboring manifolds the eigenstates $\{|j,m\rangle\}$ for different $j$ will be connected and the dynamics can be simulated efficiently by virtue of the permutation invariance \cite{shammah2018}.

\subsection{Mean-field Bloch equation}
In the following, we will restrict our discussion in the Dicke-manifold of $j=N/2$. Therefore the state in the Dicke-manifold can be represented by a Bloch vector of modulus $|\textbf{J}|=N/2$. For large $N$, within the mean-field approximation, the correlator $\langle \hat{J}_\alpha\hat{J}_\beta\rangle$ ($\alpha,\beta=x,y,z$) is factorized as $\langle \hat{J}_\alpha\hat{J}_\beta\rangle=\langle \hat{J}_\alpha\rangle\langle\hat{J}_\beta\rangle$ where $\langle\hat{A}\rangle=\text{tr}[\hat{A}\rho]$ denotes the expectation value of observable $\hat{A}$. Using the cyclic property of the trace, one can derive the equation of motion for the Bloch vector. For convenience we define $\{X,Y,Z\}=\{\langle\hat{J}_x\rangle,\langle\hat{J}_y\rangle,\langle\hat{J}_z\rangle\}/\frac{N}{2}$, thus the mean-field Bloch equations yields,
\begin{eqnarray}
\dot{X}&=&-p\frac{V}{2}YZ-(1-p)gY+\frac{\Gamma}{8}XZ,\cr\cr
\dot{Y}&=&p\left(\frac{V}{2}XZ-gZ\right)+(1-p)\left(gX-\frac{V}{2}XZ\right)+\frac{\Gamma}{8}YZ,\cr\cr
\dot{Z}&=&pgY+(1-p)\frac{V}{2}XY-\frac{\Gamma}{8}(1-Z^2),
\label{MF_BlochEq}
\end{eqnarray}
where we have adopted the constraint $X^2+Y^2+Z^2=1$ because the total angular momentum is conserved. The steady-state magnetizations of the system $\{X_\tss,Y_\tss,Z_\tss\}$ correspond to the fixed points of the system of equations (\ref{MF_BlochEq}), i.e., when $\dot{X}=\dot{Y}=\dot{Z}=0$ (the subscript `$\tss$' denotes the steady-state).

The stabilities of the fixed points are determined by the Jacobian matrix of the system of equations (\ref{MF_BlochEq}) with the elements being $M_{\alpha\beta}=\partial f_\alpha/\partial \beta$, $\alpha,\beta=X,Y,Z$ and $f_\alpha$ is the corresponding time-derivative equation in Eq. (\ref{MF_BlochEq}). Thus the Jacobian matrix reads
\begin{equation}
M = \left(
      \begin{array}{ccc}
        \frac{\Gamma Z}{8} & (p-1)g-\frac{p}{2}VZ & -\frac{p}{2}VY+\frac{\Gamma X}{8} \\
        \frac{2p-1}{2}VZ+(1-p)g & \frac{\Gamma Z}{8} & \frac{2p-1}{2}VX-pg+\frac{\Gamma Y}{8} \\
        \frac{(1-p)}{2}VY & pg+\frac{(1-p)}{2}VX & \frac{\Gamma Z}{4} \\
      \end{array}
    \right).
\label{JacobianMatrix}
\end{equation}
The appearance of an eigenvalue with positive real part when substituting the fixed points $\{X_\tss,Y_\tss,Z_\tss\}$ into Eq. (\ref{JacobianMatrix}) implies that this set of fixed points are unstable.

\section{Results}
\label{sec_results}
In this section we first present the steady-state solutions to the mean-field Bloch equations in two limit cases for $p=0$ and $p=1$. In the former case the dissipation is aligned parallel to the driving field while in the latter case the dissipation is aligned parallel to the spin interaction. In analogous to the quantum phase transition in the transverse-field Ising model, in these two limit cases, we define the disordered steady-state phase as the one that the magnetization is completely parallel to the direction of the external field, also denoted as the paramagnetic phase (PM) with zero-order parameter \cite{sachdev}. The nonzero order parameter indicates the appearance of the ordered steady-state phase or ferromagnetic (FM) phase. For instance, for the case $p=0$, the disordered steady-state PM phase is characterized by $Z_\tss=-1$ and the ordered steady-state FM phase is distinguished by the nonzero steady-state magnetization in the $x$-$y$ plane (the order parameter), namely $X_\tss\ne0$ means the ordered FM phase; while for $p=1$ the ordered FM phase is indicated by $Z_\tss\ne0$. Due to the constraint of $X^2+Y^2+Z^2=1$ the continuity (or discontinuity) of the magnetization as a function of the controlling parameter can be characterized by any of the components along $x$, $y$ and $z$ directions.

We would notice that if the external field is absent ($g/\Gamma=0$) the total Hamiltonian is reduced to the two-axis squeezing model which has been discussed with collective or independent decay in Ref. \cite{lee2014pra}. Thus in the following discussion we consider only the cases of nonzero $g$.

%As will be seen the phase transitions from the paramagnetic phase ($Z=0$) to ferromagnetic phase ($Z<0$) behave different. In particular the first-order and second-order steady-state phase transition occur in these two limit cases, respectively. We then interpolate between this two limit case by considering $0<p<1$ to figure out how does the phase transition changeover from the first to second orders. We also analyze the validity of the MF solutions by comparing to the full quantum simulation with the master equation for various number of spins.

\subsection{$p=1$}
For $p=1$, only $\hat{H}_1$ is involved in the Hamiltonian $\hat{H}$ and the dissipation is aligned parallel to the direction of the interaction.
We note that for $V=0$ the exact steady-state density matrix has been obtained by means of spin coherent states representation \cite{puri1979} and recently is shown to support the dissipative time crystal with time-translational symmetry breaking \cite{iemini2018,dosprazeres2021}.

\begin{figure}[!htbp]
\includegraphics[width=1\linewidth]{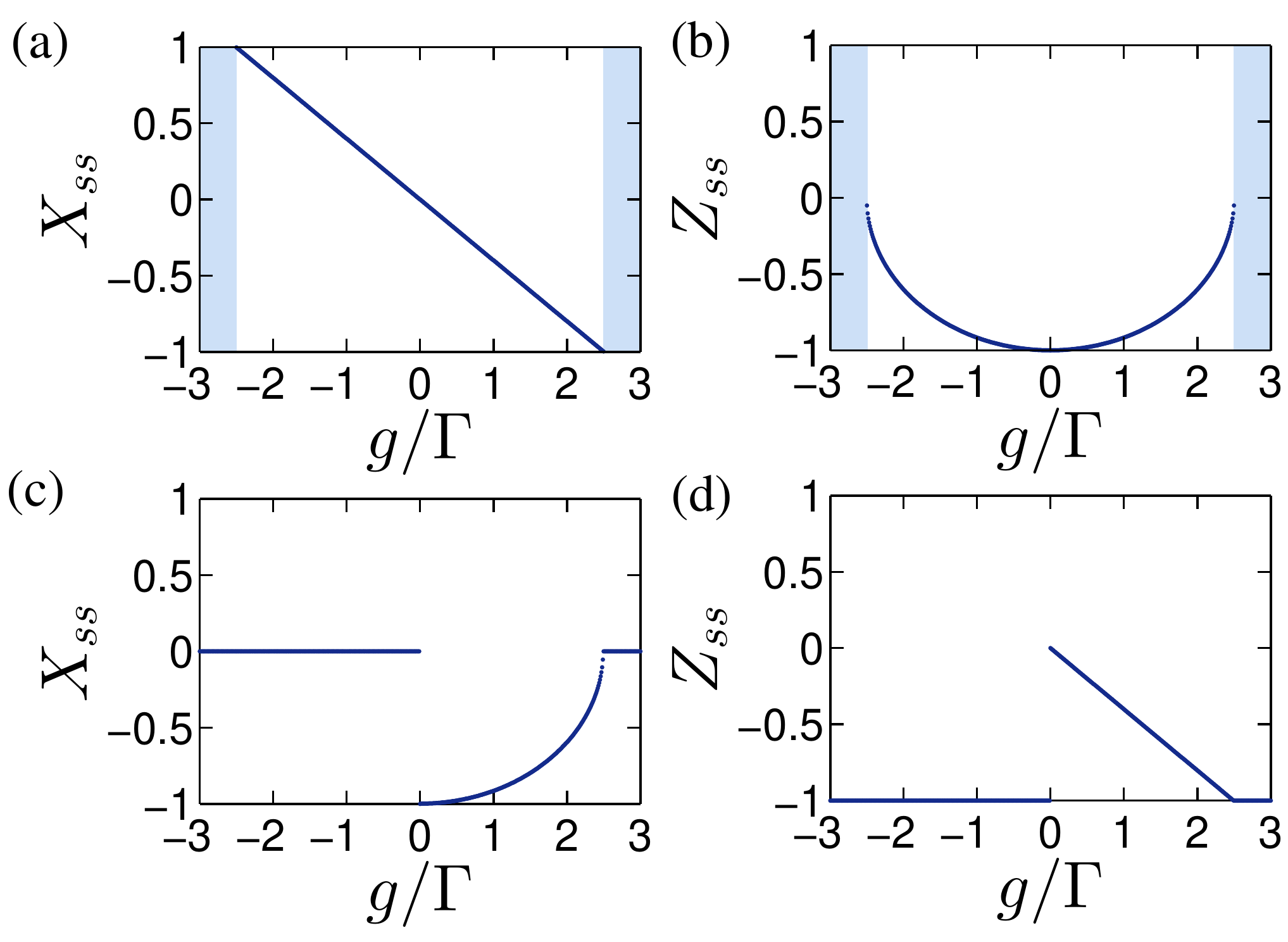}
\caption{(color online) The top panels show the steady-state values of $X$ (a) and $Z$ (b) for $p=1$.  The FM phase, indicated by the nonzero steady-state $Z$ exists in the $|g|<\sqrt{16V^2+\Gamma^2}/8$, after the continuous transition from the (shaded blue) regions where the stable fixed-points to Eq. (\ref{MF_BlochEq}) do not exist. The bottom panels show the steady-state values of $X$ (c) and $Z$ (d) for $p=0$. As $g/\Gamma$ decreases the vanishing steady-state $X$ (the order parameter) becomes nonzero at the critical point $g_+^{c}$ indicating a continuous phase transition from the PM to FM phases. As $g/\Gamma$ continues to decrease a discontinuous phase transition from the FM to PM phases occurs at $g_-^{c}$ revealed by a jump of $X$ to zero. }
\label{fig1a}
\end{figure}

\begin{figure}[!htbp]
\includegraphics[width=1\linewidth]{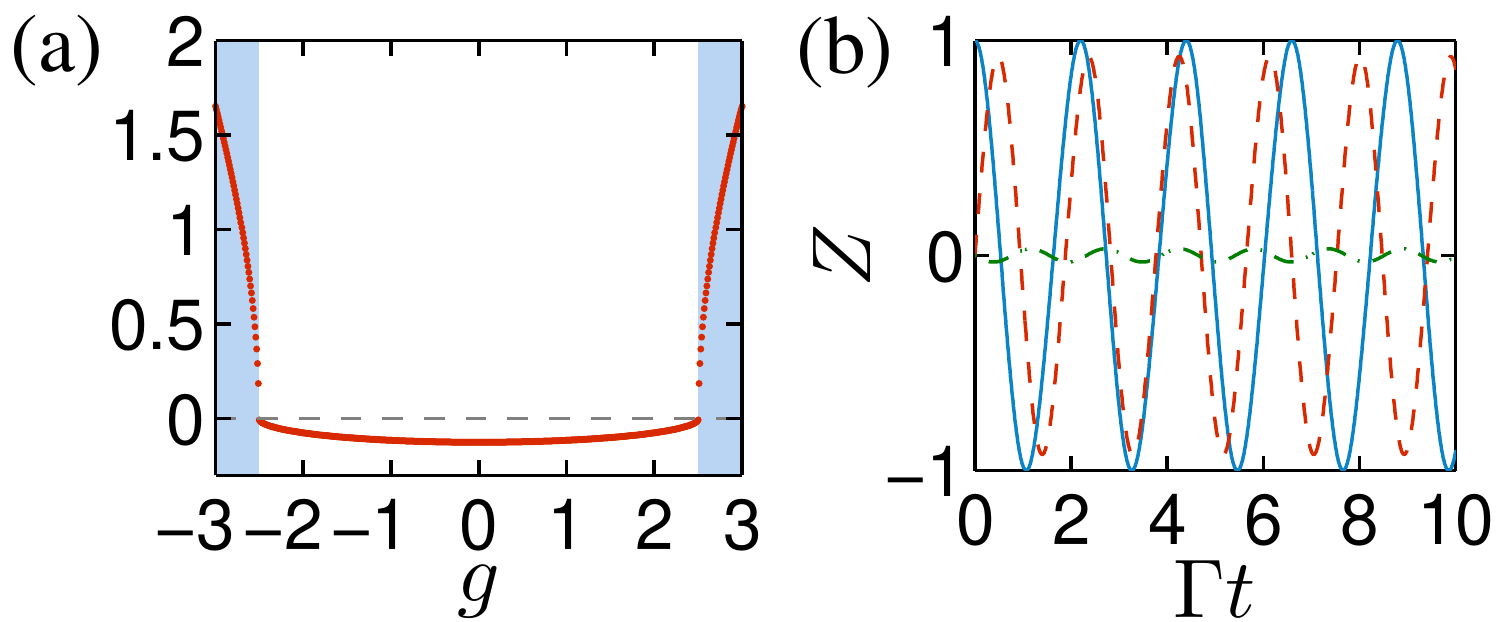}
\caption{(color online) (a) The maximal real parts of the eigenvalues of the Jacobian matrix in Eq. (\ref{JacobianMatrix}) for $p=1$. The positive maximal real parts identify the unstable regions in which there is no stable fixed points for the mean-field Bloch equations (\ref{MF_BlochEq}). (b) The time-evolution of $Z$ for various initial Bloch vectors $[X,Y,Z]=[0,0,1]$ (solid line), $[0,1,0]$ (dashed line) and $[1,0,0]$ (dotted-dashed line). The parameters are chosen as $V/\Gamma = -5$ and $g/\Gamma=3$.}
\label{fig2a}
\end{figure}

For nonzero $V$, the asymptotic steady-state magnetizations are given by the fixed points to the mean-field Bloch equations (\ref{MF_BlochEq}). However, some unstable fixed points can be eliminated by checking the signs of the real parts of the Jacobian eigenvalues, thus the stable steady-state solutions are given as follows,
\begin{eqnarray}
X_\tss&=&\frac{32gV}{16V^2+\Gamma^2},\cr\cr
Y_\tss&=&\frac{8g\Gamma}{16V^2+\Gamma^2},\cr\cr
Z_\tss&=&-\sqrt{1-\frac{64g^2}{16V^2+\Gamma^2}}.
\label{MF_sol_H1}
\end{eqnarray}
The steady-state magnetizations along the $x$ and $z$ directions for $V/\Gamma=-5$ are shown in Figs. \ref{fig1a} (a) and (b). It is obvious to see that the steady-state value of $X$ (so does $Y$, not shown) depends linearly on the Rabi frequency $g$ of the driving field. However, the $Z$-component shows a second-order phase transition from the PM ($Z_{ss}=0$ and $X_{ss}=-1$) to the FM ($Z_{ss}\ne 0$) phases. In the shaded regions of Figs. \ref{fig1a} (a) and (b) which is bound by $|g|>|V|/2$, there is no stable fixed points revealed by the positive real parts of the eigenvalues of the Jacobian (\ref{JacobianMatrix}) as shown in Fig. \ref{fig2a}(a). Moreover, in the unstable regions, the dynamics of the spin magnetization shows periodically oscillations even if in the long-time limit, however the long-time oscillations depends on the initial states as shown in Fig. \ref{fig2a}(b).

The phase diagram in the $V/\Gamma$-$g/\Gamma$ plane for $p=1$ is shown in Fig. \ref{fig3a}(c). The steady-state phase diagram is symmetric under the reflection $g\leftrightarrow -g$. The contours indicate the steady-state value of $Z_\tss$ and the nonzero $Z_\tss$ witnesses the ordered steady-state FM phase. The system enters into the FM phase at at the critical points $g_c=\pm\sqrt{16V^2+\Gamma^2}/8$ (green solid line in Fig. \ref{fig3a}(c)).

\subsubsection{The validity of the MF approximation}
\begin{figure}[!htbp]
\includegraphics[width=0.9\linewidth]{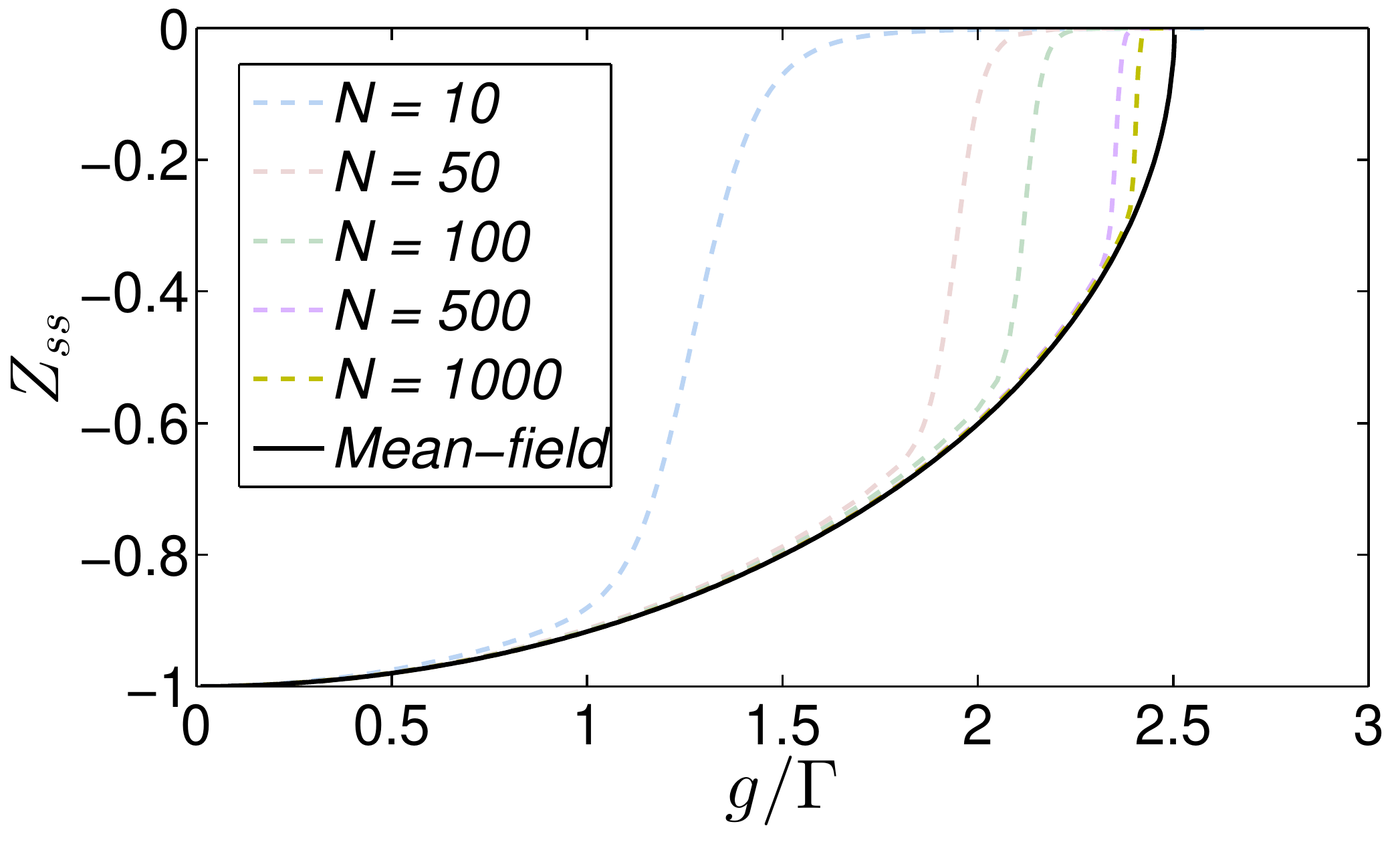}
\caption{(color online) Comparison of the steady-state values of $Z$ in finite-size systems and with mean-field approximation. The parameters are chosen as $V/\Gamma = -5$ and $p=1$. The mean-field result becomes more and more close to the finite-size computation as the number of spins $N_s$ increasing.}
\label{fig_mf}
\end{figure}

\begin{figure*}[!t]
\includegraphics[width=1\linewidth]{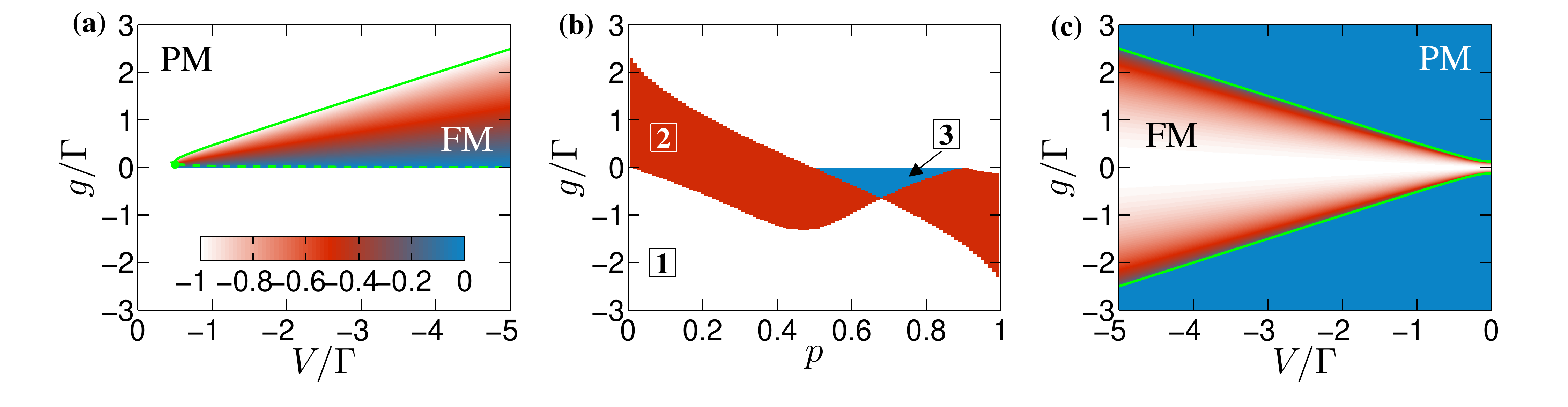}
\centering
\caption{ (color online) The phase diagram for $p=0$ (a) and $p=1$ (c) in the $g/\Gamma$-$V/\Gamma$ plane. The contour shows the steady-state values of $Z$. The green solid and dashed lines mark the phase boundaries for continuous and discontinuous phase transitions, respectively. (b) The numbers of the stable steady-state solutions to Eq. (\ref{MF_BlochEq}) in the $g/\Gamma$-$p$ plane for $V/\Gamma = -5$. In the while region there is a unique stable steady-state solution. The red and blue regions indicated the bistable and tristable regions, respectively.
\label{fig3a}
}
\end{figure*}

So far, our discussion is mainly basing on the MF approximation. The MF approximation has been proven to be accurate in capturing the steady-state properties in thermodynamic limit of high dimensional dissipative systems. In this sense the application of MF approximation in the system with all-to-all connections should be able to unveil the true behavior in the case of infinite $N$. To corroborate, we compare the steady-state magnetization of the finite-size systems and within the MF approximation. The result is shown in Fig. \ref{fig_mf}, one can see that for small $g/\Gamma$ the MF result recovers the steady-state magnetization perfectly even if for system with smaller size ($N=10$). However, for large $g/\Gamma$ the MF result gives the reliable prediction for the steady-state magnetization only for systems with larger size. The range of $g/\Gamma$ in which MF is accurate extends as the system size increasing. This tendency shows that the MF approximation is valid in the thermodynamic limit ($N\rightarrow\infty$) of the considered all-to-all model.

\subsection{$p=0$}
For $p=0$ the Hamiltonian $H_1$ is switched off while $H_0$ is switched on, the dissipation is aligned parallel to the direction of the external field. In this case, there are several steady-state solutions to Eqs. (\ref{MF_BlochEq}) given as follows,
\begin{equation}
 \{X_{\tss},Y_\tss,Z_\tss\}=\left\{\eta_{\pm,\pm},\Gamma\eta_{\pm,\pm}\xi_{\pm},8g\xi_\pm \right\},
\end{equation}
where $\xi_{\pm}=\left(2V\pm\sqrt{4V^2-\Gamma^2}\right)/\Gamma^2$ and $\eta_{\pm,\pm}=\pm\sqrt{1-(64g^2+\Gamma^2)\xi_{\pm}/4V}$.

However the constraint $X^2+Y^2+Z^2=1$ and the linear stability analysis via the Jacobian matrix reduce the number of solutions. The stable steady-state $X$ and $Z$ as functions of $g$ are shown in Figs. \ref{fig1a}(c) and (d). One can see that as $g$ decreasing the steady-state magnetization undergoes a continuous transition, the vanishing $X_{\tss}$ becomes nonzero at $g\approx 2.49$. As $g$ continues to decrease the $X_{\tss}$ shows a discontinuous transition at $g\approx 0.0063$. The continuous and discontinuous features of phase transition are in sharp contrast to the case of $p=1$ in which the FM-PM phase transitions are always continuous.

In order to have a full view we show the phase diagram in the $V/\Gamma$-$g/\Gamma$ plane for $p=0$ in Fig. \ref{fig3a}(a). The order parameter is chosen as the steady-state value $Z_\tss$: $Z_\tss=-1$ (equivalently $X_\tss=0$) means the magnetization is polarized down to the $z$ direction indicating a PM phase, otherwise there will be nonzero magnetization on the $x$-$y$ plane indicating a FM phase. One can see that there is a critical point $V^c=-\Gamma/2$, for $V\le V^c$ the ordered FM phase emerges. For fixed $V<V^c$, the FM phase is present in a closed region $g\in[g_-^c,g_+^c]$ with $g^c_{\pm}=\frac{\Gamma}{8(2V\pm\sqrt{4V^2-\Gamma^2})}$. In particular, the steady states of the system undergo a continuous phase transition at $g_+^c$ (green solid line in Fig. \ref{fig3a}(a)) while a discontinuous phase transition at $g_-^c$ (green dashed line in Fig. \ref{fig3a}(a)).

\subsection{$0<p<1$}
We have observed the continuous/discontinuous FM-PM phase transition in different cases that the dissipations are aligned parallel/vertical to the directions of the external fields. Now we are in position to investigate how the phase transition crossover from the continuous to discontinuous as the parameter $p$ varying.

For $0<p<1$ there will be multiple stable steady-state solutions to Eq. (\ref{MF_BlochEq}). In Fig. \ref{fig3a}(b) we show the number of the stable steady-state solutions in the $g/\Gamma$-$p$ plane for $V/\Gamma=-5$. Comparing with the result for $V/\Gamma=-5$ (the rightmost column) in Fig. \ref{fig3a}(a), the ordered FM phase in the case of $p=0$ becomes a bistable region as soon as $p$ being different from zero. Similarly, in the case of $p=1$, the ordered FM phase located in $g/\Gamma<0$ (the leftmost column, below zero in Fig. \ref{fig3a}(c)) becomes bistable as soon as $p$ being different from $1$. In the intermediate region $0.5\lesssim g/\Gamma \lesssim0.9$ a tristable region appears. Recall that the total angular momentum is conserved ($X^2+Y^2+Z^2=1$), the continuity of the behavior of magnetizations can be witnessed by any component. Here we choose the $z$ component to characterize the possible phase transition.

\begin{figure}[!htbp]
\includegraphics[width=0.9\linewidth]{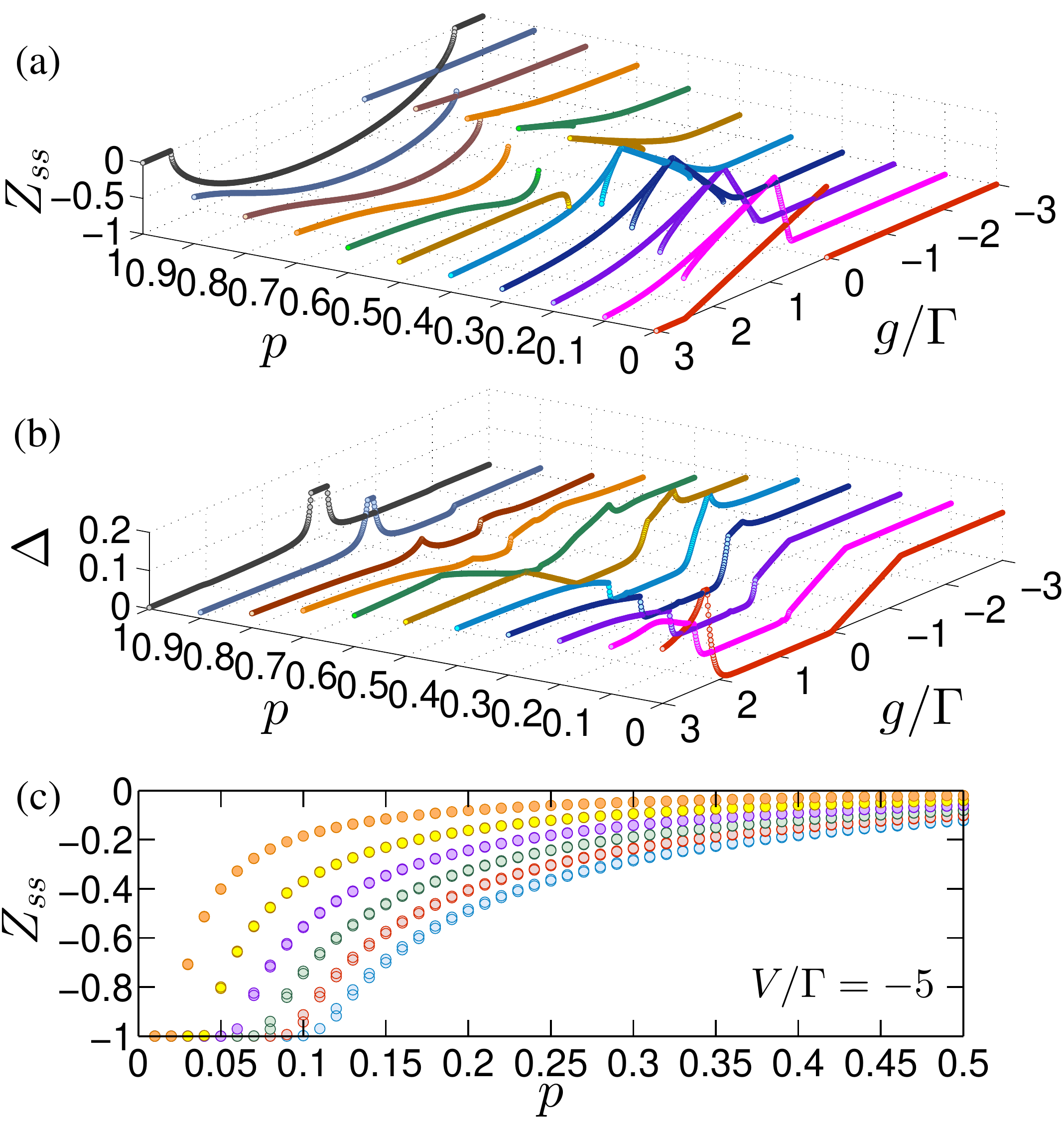}
\caption{(color online) (a) The steady-state values of $Z$ as functions of $g/\Gamma$ for various $p$. Each set of data (in the same color) displays the stable steady-state $Z$ corresponding to a horizontal cut along a fixed $g/\Gamma$ in Fig. \ref{fig3a}(b). (b) The Liouvillian gap of finite-size system as a function of $g/\Gamma$ for various $p$ for $N=50$. (c) The stable steady-state $Z$ as a function of $p$ for different values of $g/\Gamma$. From the right to left the data are produced with $g/\Gamma=\{-0.55-0.45,-0.35,-0.25,-0.15,-0.05\}$.
The parameter is chosen as $V/\Gamma = -5$.}
\label{fig4b}
\end{figure}

In Fig. \ref{fig4b}(a) we show all the stable steady-state $Z$ as functions of $g/\Gamma$ for various $p$. One can see that, starting from $p=0$, for large amplitude $g/\Gamma$, the value of $Z_\tss$ monotonically increases from $-1$ to $0$ meaning that all the spins gradually change their orientation from $z$-direction to $x$-direction. During the whole process there is always a unique steady state for each $p$ and the crossover of the spin alignment is smooth without any transition. However, another stable branch is observed for $0<g/\Gamma\lesssim 2.5$ and $p\lesssim 0.5$. Such branch appears as soon as $p$ being different from zero and become more visible as $p$ increasing which is responsible for the bistable region in Fig. \ref{fig3a}(b) for positive $g/\Gamma$.

\jj{The crossover from discontinuous to continuous phase transition can also be understood via the Liouvillian spectral theory. The matrix form of the Liouvillian superoperator in Eq. (\ref{QME}) is constructed as follows,
\begin{equation}
\mathbb{L} = -i(\mathbb{I}\otimes \hat{H}^T-\hat{H}\otimes \mathbb{I})+\frac{\Gamma}{2N}(2\hat{J}_-\otimes\hat{J}_--\mathbb{I}\otimes\hat{J}_+\hat{J}_--\hat{J}_+\hat{J}_-\otimes\mathbb{I}),
\end{equation}
where the superscript $T$ denotes the transpose of matrix \cite{arg2022}. The real parts of the eigenvalues of $\mathbb{L}$ are always non-positive while the eigenvector associated to the zero eigenvalue corresponds to the steady state. The Liouvillian gap $\Delta$ is defined by the nonzero eigenvalue $\lambda_1$ with the largest negative real part, i.e. $\Delta=|\text{Re}(\lambda_1)|$, which is also called the asymptotic decay rate since it
describes the slowest relaxation time scale toward the steady state \cite{kessler2012}. The closure of the Liouvillian gap implies the appearance of ordered phase or bistable region \cite{minganti2018}. In Fig. \ref{fig4b}(b), we show the Liouvillian gap of a finite-size system with $N=50$. For $p=0$ the ordered FM phase is marked by the closing of the Liouvillian gap. As $p$ increasing, the closure of the Liouvillian gap that indicates the bistability of the steady state shifts in accordance with the steady-state values $Z_{ss}$ shown in Fig. \ref{fig4b}(a). Moreover as $p$ increasing the critical point given by the smaller $g/\Gamma$ shifts to $g\approx -2.49$ which is the critical point for $p=1$. We emphasize that since a finite-size system is considered in Fig. \ref{fig4b}(b), the deviation of the steady-state phases revealed by the Liouvillian spectra and by the solution to the mean-field Bloch equation is visible, for example, the bumps are observed around $g=0$ for $p\gtrsim 0.7$. Such singularities are expected to be suppressed in the system with larger $N$.
}

It is interesting that for $g/\Gamma<0$ although the system always undergoes a transition from a phase with unique steady state to bistability as $p$ increases, such transition becomes more likely to a second-order PM-FM phase transition as $g/\Gamma$ approaching to zero. We highlight the steady-state $Z$ as a function of $p$ for various $g/\Gamma$ in Fig. \ref{fig4b}(c). For smaller $g/\Gamma$ ($\le -0.25$) it is obvious to see that there are two branches of steady-state $Z$. Moreover these two branches becomes more and more close to each other as $g/\Gamma$  approaches to zero and eventually merges.

\begin{figure}[!htbp]
\includegraphics[width=0.9\linewidth]{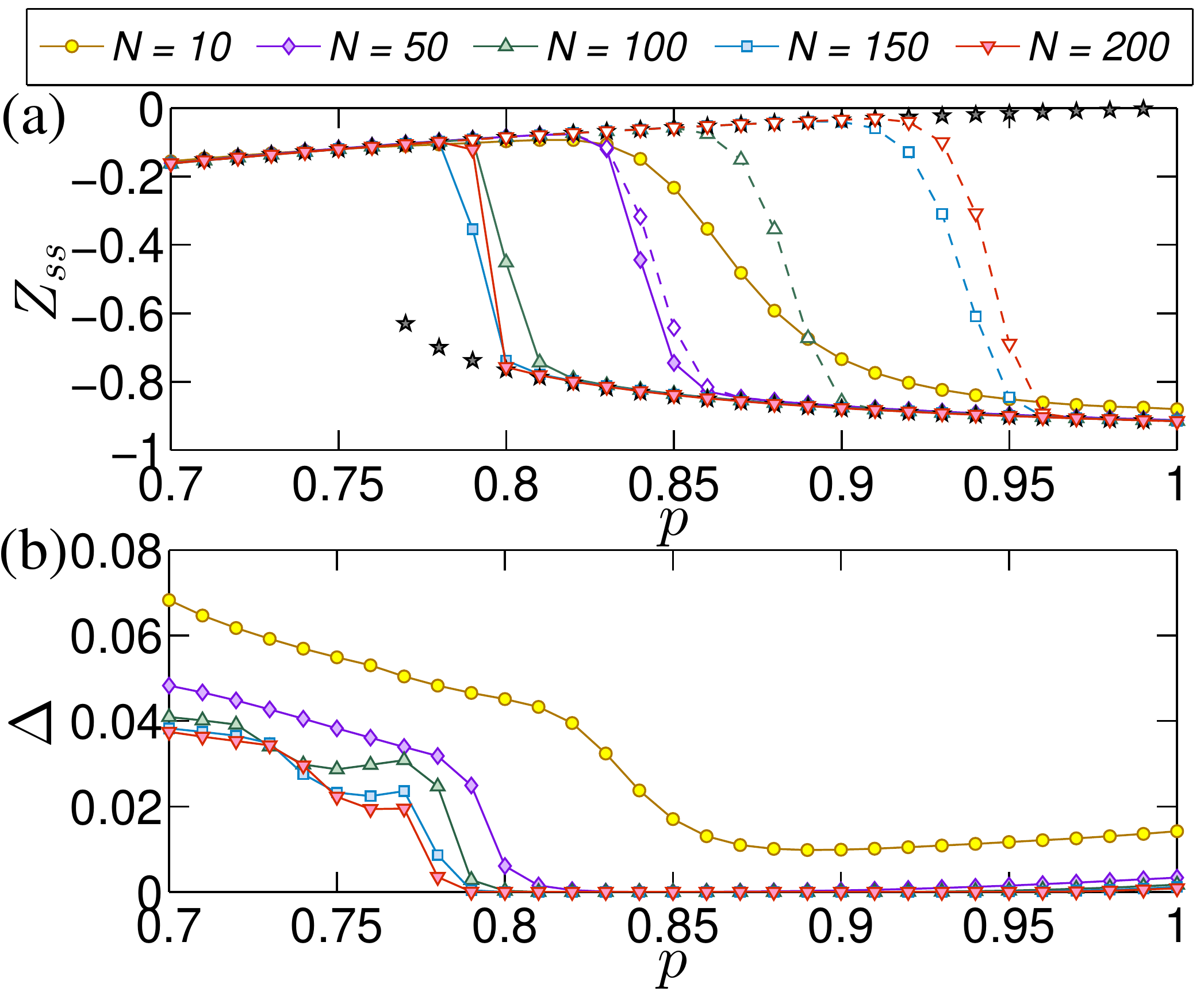}
\caption{(color online) (a) The Steady-state magnetization as a function of the ratio $p$. The black pentagram denotes the mean-field solutions for infinite $N$. The filled and empty symbols indicate the two branches of steady-state solutions for various system size $N$ when sweep from larger to smaller $p$ (solid) and conversely (dashed lines). (b) The Liouvillian gap $\Delta$ as a function of $p$ for various system size $N$ as indicated in the legend. Other parameters are chosen as $V/\Gamma = -5$ and $g/\Gamma = -1$. }
\label{fig5a}
\end{figure}

The bistable region for the relative large $p$ in Fig. \ref{fig3a}(b) is a consequence of the extension of the bistable region in the finite-size system. In Fig. \ref{fig5a}(a) we show the steady-state $Z$ as a function of $p$ in a seris of finite-size systems. One can see that for small system, e.g. $N=10$ the steady-state magnetization is unique and it shows a crossover as $p$ varying. As the system size keep enlarging, the steady-state magnetization is not unique and a bistable behavior emerges. The two branches correspond to the solutions obtained when sweeping the from larger to smaller value of $g$, or conversely. Moreover the bistable region extends as the system size increasing and is expected to converge in the infinite size as predicted by the mean-field approximation.

%The emergence of bistable region in large-size system is revealed by the Liouvillian spectrum analysis. The matrix form of the Liouvillian superoperator in Eq. (\ref{QME}) is constructed as follows,
%\begin{equation}
%\mathbb{L} = -i(\mathbb{I}\otimes \hat{H}^T-\hat{H}\otimes \mathbb{I})+\frac{\Gamma}{2N}(2\hat{J}^-\otimes\hat{J}^--\mathbb{I}\otimes\hat{J}^+\hat{J}^--\hat{J}^+\hat{J}^-\otimes\mathbb{I}),
%\end{equation}
%where the superscript $T$ denotes the transpose of matrix. The real parts of the eigenvalues of $\mathbb{L}$ are always negative. The eigenvector associated to the zero eigenvalue corresponds to the steady state. The Liouvillian gap $\Delta$ is defined by the nonzero eigenvalue $\lambda_1$ with the largest negative real part, i.e. $\Delta=|\text{Re}(\lambda_1)|$, which is also called the asymptotic decay rate since it describes the slowest relaxation scale toward the steady state \cite{kessler2012}. The closure of the Liouvillian gap implies the appearance of ordered phase or bistable region \cite{minganti2018}.

In Fig. \ref{fig5a}(b) we show the Liouvillian gap as a function of $p$ for the systems of different sizes. For each $N$, the Liouvillian gap is closed over an intermediate region of $p$ indicating the existence of the degenerated steady states. The gapless region extends as the system size increasing. This can be observed on the one hand as $N$ increasing the critical point of $p$  shifts right towards to $p\approx0.77$ which is predicted by the mean-field Bloch equations in thermodynamic limit, on the other hand $p\gtrsim 0.9$ the Liouvillian gap $\Delta$ tends to be closed as $N$ increasing. The tendency from hysteresis to bistable behavior of the steady state as $N$ goes to infinite in the considered model is in sharply contrast to the dissipative transverse-field Ising model with nearest-neighboring interactions \cite{jin2018}. In the latter, the bistable region can be observed by means of cluster mean-field treatment but it shrinks as the short-range correlations are gradually included and eventually replaced by a discontinuous phase transition. It is the long-range correlations in the all-to-all model guarantees the validity of mean-field approximation.

\section{Summary}
\label{sec_summary}
In summary, we have investigated the steady-state phase transitions in a mixture of two all-to-all transverse-field Ising models with collective decay. The dissipation tends to incoherently flip the spins down to the $z$-direction. The considered Hamiltonian is composed of two ingredients $\hat{H}_1$ and $\hat{H}_2$ as explained in Eq. (\ref{Ham}). The ratio of the two Hamiltonian is controlled by the parameter $p$. By means of mean-field approximation and linear stability analysis, we have shown that the only continuous steady-state phase transition is present in the case of $p=1$, namely, when the orientation of the external field is perpendicular to the dissipation. While discontinuous steady-state phase transition can be occur in the case of $p=0$, that is the orientation of the external field is parallel to the dissipation.

We have investigated how the orders of steady-state phase transitions crossover as the ratio of $\hat{H}_1$ and $\hat{H}_2$ varying. Due to the competition of the two components in the total Hamiltonian, we found that there are multiple stable steady-state solutions to the mean-field Bloch equation for $0< p<1$. In particular, we fixed the Rabi frequency of the external field and tracked the orientation of the steady-state magnetization when the ratio $p$ is tuned from $0$ and $1$. Although the master equation does not possess any symmetry due to the mixture of two ingredients in the Hamiltonian $\hat{H}$, the conservation of the total angular momentum enables us to characterize the continuity (or discontinuity) of the behavior of steady-state magnetization through one of the spin component.

We found that a quasi continuous PM-FM transition and the nonanalyticity of the steady-state $Z$ is due to the merging of two branches of stable solutions. Similar behaviors has been observed in the steady-state magnetization of a multiply-spin system interacting with a common cavity mode \cite{hannukainen2018}. The emergence of the bistability as $\hat{H}_1$ becomes more dominant is a consequence of the gradually extension of the hysteresis in finite-size systems. This is in contrast to the dissipative transverse-field Ising model with nearest-neighboring interaction in which the bistability in the small size cluster mean-field approximation is replaced by the discontinuous phase transition in thermodynamic limit \cite{jin2018}. The existence of the bistability is also revealed by the gapless Liouvillian spectrum in the bistable region.

\jj{
It should be pointed that our calculation is restricted on the Dicke-manifold of  $j=N/2$.  Since the total angular momentum is conserved during the time-evolution generated by the collective dissipation, the present results remain the same (with the normalized Bloch vector) if one works in other subspaces spanned by $\{|j,m\rangle\}$ where $j<N/2$ and $|m|\le j$. Moreover, because the collective decay does not connect the neighboring Dicke ladders, our analysis also applies to the case of incoherent mixture of the Dicke-manifold with different angular momentum quantum numbers.
When more than two Dicke-manifolds are involved, for example, the system is initialized in a superposition of states belong to different Dicke-manifolds, the coherence between the involved subspaces may produce nonzero off-block-diagonal elements of the joint density matrix. For such case the permutation invariance of the considered model can be used to solve the full quantum master equation \cite{shammah2018}.
}

Finally, we note that the explicit symmetry is absent when both $\hat{H}_0$ and $\hat{H}_1$ are involved in the considered model. Therefore the phase transitions appearing in the case of $0<p<1$ are not accompanied by any spontaneous symmetry breaking. Recently, such anomalous phase transition without symmetry breaking in a nonequilibrium open quantum system has been investigated within the Liouvilian spectrum theory. It is proven that the spontaneous symmetry breaking is not a necessary condition for the occurrence of the continuous dissipative phase transition \cite{minganti2021njp}. For the future work, we expect to gain insightful understanding for the crossover from discontinuous to continuous dissipative phase transition in the framework of spectral theory of Liouvillians. Experimentally, the manipulating and tuning the interactions among among many-body system in the scalable quantum simulators through the use of Rydberg atoms is fast developed, it provides exciting opportunity to test these predictions in the laboratory \cite{carr2013,gutierrez2017,ding2020,bernien2017,lienhard2018,keesling2019,nill2022}.

\section*{ACKNOWLEDGMENTS}
This work is supported by National Natural Science Foundation of China under Grant No. 11975064.

\end{document}